\documentclass[sigconf]{acmart}
\usepackage{algorithm}
\usepackage[noend]{algpseudocode}
\usepackage{cite}
\usepackage{amsmath,amsfonts}
\usepackage{adjustbox}
\usepackage{multirow}
\usepackage{multicol}
\usepackage{graphicx}
\usepackage{textcomp}
\usepackage{xcolor}
\def\BibTeX{{\rm B\kern-.05em{\sc i\kern-.025em b}\kern-.08em
    T\kern-.1667em\lower.7ex\hbox{E}\kern-.125emX}}

\usepackage{mathtools}

\usepackage{tcolorbox}

\usepackage{listings}
\usepackage{subcaption}

\usepackage{xcolor}

\definecolor{codegreen}{rgb}{0,0.6,0}
\definecolor{codegray}{rgb}{0.5,0.5,0.5}
\definecolor{codepurple}{rgb}{0.58,0,0.82}
\definecolor{backcolour}{rgb}{0.95,0.95,0.92}
\definecolor{cellgreen}{HTML}{009901}
\definecolor{cellyellow}{HTML}{FFF700}
\definecolor{cellorange}{HTML}{FF6F00}
\definecolor{clgreen}{HTML}{34FF34}
\definecolor{clyellow}{HTML}{FDFF7E}
\definecolor{clorange}{HTML}{FE996B}
\definecolor{cellud}{HTML}{EE442F}
\definecolor{cellue}{HTML}{63ACBE}
\definecolor{celluo}{HTML}{CCBE9F}

\definecolor{cellgray}{HTML}{aeaeae}

\lstdefinestyle{mystyle}{
    backgroundcolor=\color{backcolour},   
    commentstyle=\color{codegreen},
    keywordstyle=\color{blue},
    numberstyle=\tiny\color{codegray},
    stringstyle=\color{codepurple},
    basicstyle=\ttfamily\tiny,
    breakatwhitespace=false,         
    breaklines=true,                 
    captionpos=b,                    
    keepspaces=true,                 
    numbers=left,                    
    numbersep=3pt,                  
    showspaces=false,                
    showstringspaces=false,
    showtabs=false,                  
    tabsize=2
}
 
\newcommand\SBST[1]{{\textcolor{black}{SBST guided by DP}}}
\newcommand\SBSTLong[1]{{\textcolor{black}{Defect prediction guided SBST}}}

\usepackage{tikz}
\usetikzlibrary{arrows.meta}
\usetikzlibrary{arrows,shapes,decorations,automata,backgrounds,petri}
\tikzset{point/.style = {fill=black,circle,inner sep=0.7pt}}
\tikzset{
  graph vertex/.style={
    circle,
    draw,
  },
    graph vertexx/.style={
    draw,
  },
  graph directed edge/.style={
    ->,
    >=stealth,
    thick,
  },
  graph tree edge/.style={
    graph directed edge
  },
  graph forward edge/.style={
    graph directed edge,
    every edge/.style={
      edge node={node [fill=white,font=\scriptsize] {f}},
      loosely dotted,
      draw,
    },
  },
  graph back edge/.style={
    graph directed edge,
    every edge/.style={
      edge node={node [fill=white,font=\scriptsize] {b}},
      densely dotted,
      draw,
    },
  },
  graph cross edge/.style={
    graph directed edge,
    every edge/.style={
      edge node={node [fill=white,font=\scriptsize] {c}},
      dotted,
      draw,
    },
  },
}

\usepackage{scalerel}
\usetikzlibrary{svg.path}

\definecolor{orcidlogocol}{HTML}{A6CE39}
\tikzset{
  orcidlogo/.pic={
    \fill[orcidlogocol] svg{M256,128c0,70.7-57.3,128-128,128C57.3,256,0,198.7,0,128C0,57.3,57.3,0,128,0C198.7,0,256,57.3,256,128z};
    \fill[white] svg{M86.3,186.2H70.9V79.1h15.4v48.4V186.2z}
                 svg{M108.9,79.1h41.6c39.6,0,57,28.3,57,53.6c0,27.5-21.5,53.6-56.8,53.6h-41.8V79.1z M124.3,172.4h24.5c34.9,0,42.9-26.5,42.9-39.7c0-21.5-13.7-39.7-43.7-39.7h-23.7V172.4z}
                 svg{M88.7,56.8c0,5.5-4.5,10.1-10.1,10.1c-5.6,0-10.1-4.6-10.1-10.1c0-5.6,4.5-10.1,10.1-10.1C84.2,46.7,88.7,51.3,88.7,56.8z};
  }
}

\newcommand\orcidicon[1]{\href{https://orcid.org/#1}{\mbox{\scalerel*{
\begin{tikzpicture}[yscale=-1,transform shape]
\pic{orcidlogo};
\end{tikzpicture}
}{|}}}}

\usepackage{hyperref}

\title{How good does a Defect Predictor need to be to guide Search-Based Software Testing?}

\author{Anjana Perera}
\email{Anjana.Perera@monash.edu}
\orcid{0000-0002-5080-9276}
\affiliation{%
  \institution{Faculty of Information Technology}
  \institution{Monash University}
  \city{Melbourne}
  \country{Australia}
}

\author{Burak Turhan}
\email{Burak.Turhan@oulu.fi}
\orcid{0000-0003-1511-2163}
\affiliation{%
  \institution{Faculty of Information Technology and Electrical Engineering}
  \institution{University of Oulu}
  \city{Oulu}
  \country{Finland}
}

\author{Aldeida Aleti}
\email{Aldeida.Aleti@monash.edu}
\orcid{0000-0002-1716-690X}
\affiliation{%
  \institution{Faculty of Information Technology}
  \institution{Monash University}
  \city{Melbourne}
  \country{Australia}
}

\author{Marcel B\"{o}hme}
\email{marcel.boehme@acm.org}
\orcid{0000-0002-4470-1824}
\affiliation{%
  \institution{Monash University, Australia}
}
\affiliation{%
  \institution{MPI-SP, Germany}
}

\begin{document}

\begin{abstract}

Defect predictors, static bug detectors and humans inspecting the code can locate the parts of the program that are buggy before they are discovered through testing. 
Automated test generators such as search-based software testing (SBST) techniques can use this information to direct their search for test cases to likely buggy code, thus speeding up the process of detecting existing bugs. 
However, often the predictions given by these tools or humans are imprecise, which can misguide the SBST technique and may deteriorate its performance. 
In this paper, we study the impact of imprecision in defect prediction on the bug detection effectiveness of SBST.

Our study finds that the recall of the defect predictor, i.e., the probability of correctly identifying buggy code, has a significant impact on bug detection effectiveness of SBST with a large effect size. On the other hand, the effect of precision, a measure for false alarms, is not of meaningful practical significance as indicated by a very small effect size. 
In particular, the SBST technique finds 7.5 less bugs on average (out of 420 bugs) for every 5\% decrements of the recall.

In the context of combining defect prediction and SBST, our recommendation for practice is to increase the recall of defect predictors at the expense of precision, while maintaining a precision of at least 75\%. To account for the imprecision of defect predictors, in particular low recall values, SBST techniques should be designed to search for test cases that also cover the predicted non-buggy parts of the program, while prioritising the parts that have been predicted as buggy.

\end{abstract}

\maketitle

\section{Introduction}

Search-based software testing (SBST) techniques search for test cases to optimise a given coverage criterion such as branch coverage, method coverage, or a combination of the two. Coverage related heuristics, such as branch distance~\citep{korel1990automated, mcminn2011search} and approach level~\citep{panichella2017automated} are used to guide the search for test cases to cover the uncovered areas in the program. SBST techniques are known to be effective at achieving high code coverage~\citep{panichella2017automated, panichella2018large}. While it is necessary for a test case to cover the buggy code to find a bug, just covering the buggy code may not be sufficient to discover the bug~\citep{shamshiri2015automatically, perera2020defect}. In fact, SBST techniques guided only by coverage have been shown to struggle in terms of bug detection~\citep{shamshiri2015automatically, perera2020defect, almasi2017industrial, salahirad2019choosing}. 
This is because the SBST techniques have no guidance in terms of where the buggy code is likely to be located, and hence spend most of the search effort in non-buggy code which constitutes a greater portion of the code base.

Defect predictors~\citep{hall2011systematic} and static bug detectors~\citep{ayewah2008using} can estimate the locations of the bugs effectively. 
Most of the defect predictors use classifiers trained on an existing dataset with features related to various metrics like code size, code complexity, change history, etc., and whether the components (e.g., file/class or method) are buggy or not~\citep{giger2012method}. 
Static bug detectors statically check the code against pre-defined bug patterns and label the buggy code (e.g., line) with a warning~\citep{habib2018many}. 
Both defect predictors and static bug detectors are used in the industry to assist developers in manual code reviews~\citep{lewis2013does, googledefect, aftandilian2012building, sadowski2015tricorder}. 
Defect predictors have also been used to inform automated testing techniques; G-clef~\citep{paterson2019empirical} is a test case prioritisation strategy that prioritises test cases that cover highly likely to be defective classes, and SBST$_{DPG}$~\citep{perera2020defect} and BTG~\citep{hershkovich2019prediction} are time budget allocation techniques which allocate a higher time budget to highly likely to be defective classes.

Often, the predictions produced by defect predictors are not perfectly accurate. The false positives and false negatives can significantly hinder the potential benefits of these tools. For example, false positives (i.e., wrongly labelling a program as buggy) result in SBST techniques looking for bugs in non-buggy areas in code, thus spending valuable search resources in vain. On the other hand, false negatives (i.e., labelling a buggy program as non-buggy) can result in SBST techniques not generating tests for buggy areas in code. Previous work that use defect predictors to guide SBST techniques report on improved bug detection performance of SBST~\citep{perera2020defect, hershkovich2019prediction}. The defect predictors used in these approaches have a relatively high performance, e.g., the defect predictor used by Perera et al.~\citep{perera2020defect} had a recall of 85\%, and Hershkovich et al.~\citep{hershkovich2019prediction} employed a defect predictor which had an area under curve (AUC) of 0.95. 

The performance of defect predictors, however can vary, e.g., from as low as 5\% and 25\% to as high as 95\% and 85\% for precision and recall, respectively~\citep{hall2011systematic}. 
Given such wavering performance, the question that we address in this paper is \textit{``What is the impact of imprecise predictions on the bug detection performance of SBST?''}. 

To answer this question, we simulate defect predictors for different value combinations of recall and precision in the range 75\% and 100\% (Section~\ref{subsec:dp_sim}). 
Defect predictors having recall and precision above 75\% are considered acceptable defect predictors~\citep{zimmermann2009cross}. 
We employ the state of the art DynaMOSA~\citep{panichella2017automated} as the SBST technique which is guided by the defect predictions (DP) (see Section~\ref{subsec:sbst_dpg}), which we refer to as \textit{\SBST{}} throughout the paper. 
We evaluate how the bug detection effectiveness of \SBST{} changes with the different levels of imprecision when applied to 420 bugs from the Defects4J dataset~\citep{defects4jweb} (Section~\ref{subsec:exp_subjects}).

The results from our experimental evaluation reveal that the recall of the defect predictor has a significant impact on the bug detection effectiveness of SBST with a large effect size. 
More specifically, \SBST{} finds 7.5 less bugs on average (out of 420 bugs) for every 5\% decrements of recall. 
On the other hand, the impact of precision is not of practical significance as indicated by a very small effect size, hence we conclude that the precision of defect predictors has negligible impact on the bug detection effectiveness of SBST, as long as one uses a defect predictor with acceptable performance, i.e., with precision and recall greater than 75\%. 
Further analysis into the results reveals that the impact of recall is greater for the bugs that are isolated in one method than for the bugs that are spread across multiple methods.

In summary, the contribution of this work is a comprehensive experimental analysis of the impact of imprecision of defect predictions on bug detection effectiveness of SBST. 
The experimental evaluation involving 420 bugs from 6 open source Java projects took roughly 180,750 CPU-hours in total. 
Based on the results of our study we make the following recommendations; 
\vspace{-3mm}
\begin{enumerate}
    \item SBST techniques must take potential errors in the predictions into account, in particular the false negatives. One possible solution is to prioritise predicted buggy parts of the program, while guiding the search with a certain probability towards locations that are predicted as not buggy. 
    \item In the context of combining defect prediction and SBST, it is beneficial to increase the recall of the defect predictor by sacrificing precision, while maintaining the precision above 75\%. One potential solution is to lower the cut-off point of the classifier such that more components will be labelled as buggy at the expense of more false positives. 
\end{enumerate}
The source code of \SBST{}, defect predictor simulator, post processing scripts and data are publicly available in the following link: https://figshare.com/s/a8d75f161b8cfa11d297

\vspace{-2mm}
\section{Methodology}

Our aim is to understand how the defect prediction imprecision impacts the bug finding performance of SBST. To this end, we design a study that addresses the following research question: 

\vspace{3mm}

\begin{center}
    \textit{RQ: What is the impact of the imprecision of defect prediction on bug detection performance of SBST?}
\end{center}

\vspace{3mm}

To address this research question, we measure the effectiveness of SBST in terms of finding bugs when using defect predictors with different levels of imprecision. We use DynaMOSA~\citep{panichella2017automated}, a state-of-the-art SBST technique, and incorporate predictions about buggy methods in order to guide the search for test cases towards likely buggy methods (see Section~\ref{subsec:sbst_dpg}), which we refer as \textit{\SBST{}} throughout the paper. 
Fine-grained defect predictions such as method level is chosen so that the location of the bug is narrowed down better than coarse-grained defect predictions such as class level. 
Hence the defect predictors at method level provide additional information to the SBST technique such that it can further narrow down the search for test cases to likely buggy methods.

We measure defect predictor imprecision using recall and precision. 
Recall and precision have been widely used in previous work to report the performance of defect predictors~\citep{hall2011systematic, hosseini2017systematic}.
We consider a defect predictor with either recall or precision less than 75\% is not an acceptable defect predictor, as recommended by Zimmermann et al.~\citep{zimmermann2009cross}.
Hence, we simulate defect predictors for varying levels of recall and precision in the range 75\% to 100\% (see Section~\ref{subsec:dp_sim}) and measure the impact on the bug detection performance of SBST by the prediction imprecision.

\vspace{-2mm}
\subsection{Defect Prediction Simulation}
\label{subsec:dp_sim}

To measure the bug detection performance of SBST against the imprecision of defect predictions, we simulate defect predictor outcomes at various levels of performance in the range 75\% and 100\% for both precision and recall. 
Recall is the rate of the defect predictor identifying buggy methods. It is calculated as in Equation.~\eqref{eq:recall}, where $tp$ is the number of true positives, i.e., number of buggy methods that are correctly classified, and $fn$ is the number of false negatives, i.e., number of buggy methods that are incorrectly classified.

\begin{equation}\label{eq:recall}
    \text{recall} = \frac{tp}{tp + fn}    
\end{equation}

Precision is the rate of the correct buggy methods labelled by the defect predictor. 
It can be calculated as in Equation~\eqref{eq:precision}, where $fp$ is the number of false positives, i.e., number of non-buggy methods that are incorrectly classified as buggy methods.

\begin{equation}\label{eq:precision}
    \text{precision} = \frac{tp}{tp + fp}    
\end{equation}

We simulate defect predictions from 75\% to 100\% recall in 5\% steps, with 75\% and 100\% precision. Thus, there are altogether 12 defect predictor configurations, with the following values of (precision, recall): (75\%, 75\%), (75\%, 80\%), (75\%, 85\%), (75\%, 90\%), (75\%, 95\%), (75, 100\%), (100\%, 75\%), (100\%, 80\%), (100\%, 85\%), (100\%, 90\%), (100\%, 95\%), (100, 100\%). 
Our preliminary experiments suggest that the bug detection performance of \SBST{} changes by a small margin when the precision is changed from 100\% to 75\%, while keeping the recall unchanged. 
On the other hand, the bug detection performance of \SBST{} changes by a large margin when only the recall is changed from 100\% to 75\%. 
Hence, we decide to consider only the values of 75\% and 100\% for precision, while recall is sampled at 5\% steps.

The output of the simulated defect predictor is binary, i.e., method is buggy or not buggy, similar to most of the existing defect predictors. 
Some of the existing defect predictors output the likelihood of the components being buggy or the ranking of the components according to their likelihood of being buggy. 
Since we employ a theoretical defect predictor and not a specific one, we resort to the generic defect predictor, which is the one that gives a binary classification. 

\begin{algorithm}[!ht]
\caption{Defect Predictor Simulation}\label{algo:dp_sim}
\hspace*{\algorithmicindent} \textbf{Input:} \hspace*{\algorithmicindent} $r$, \hspace*{\algorithmicindent} $p$ \Comment{recall and precision}\\
\hspace*{\algorithmicindent} $M = \{m_{1}, \ldots, m_{k}\}$ \Comment{ground truth}
\begin{algorithmic}[1]
\Procedure{SimulateDefectPredictor}{}
\State $d \gets$ \Call{Count}{$m_i$} for $m_i \in M$ s.t. $m_i$ = 1 \label{algoline:d_count} 
\State $nd \gets |M| - d$ \label{algoline:nd_count} 
\State $M_b \gets$ $\{i \, | \, \forall i \in [1, k] \wedge  m_i = 1\}$ \label{algoline:buggy_indices} 
\State $M_n \gets$ $\{i \, | \, \forall i \in [1, k] \wedge m_i = 0\}$ \label{algoline:non_buggy_indices} 
\State  $tp \gets d * r$ \label{algoline:tp_calc} 
\State $fp \gets tp * (1 - p) / p$ \label{algoline:fp_calc} 
\State $C_b \gets$ \Call{RandomChoice}{$M_b, tp$} $\cup$ \Call{RandomChoice}{$M_n, fp$} \label{algoline:buggy_method_classfic} 
\State $C \gets$ $\{c_i = 1\, | \, \forall i \in [1, k] \wedge i \in C_b$, $c_i = 0 \, | \, \forall i \in [1, k] \wedge i \notin C_b\}$ \label{algoline:classifications}
\State \Call{Return}{$C$}
\EndProcedure
\end{algorithmic}
\end{algorithm}

Algorithm~\ref{algo:dp_sim} illustrates the steps of simulating the defect predictor outputs for a given recall and precision combination. 
The procedure \Call{SimulateDefectPredictor}{} receives the set of methods in the project with  the ground truth labels for their defectiveness, $M = \{m_1,\ldots,m_k\}$, where 
$$
  m_i= \begin{cases}
    1       & \quad \text{if method with index } i \text{ is buggy}\\
   0  & \quad \text{otherwise }
  \end{cases}
$$ 
and outputs a set of labels for each method in the project, $C = \{c_1,\ldots,c_k\}$, where 
$$
  c_i= \begin{cases}
    1       & \quad \text{if method with index } i \text{ is predicted buggy}\\
   0  & \quad \text{otherwise }
  \end{cases}
$$ 

First, it calculates the number of buggy ($d$) and non-buggy methods ($nd$) in the project (lines~\ref{algoline:d_count}-\ref{algoline:nd_count} in Algorithm~\ref{algo:dp_sim}). Next, it finds the set of indices of all the buggy ($M_b$) and non-buggy methods ($M_n$) in the project (lines~\ref{algoline:buggy_indices}-\ref{algoline:non_buggy_indices}). The true positives ($tp$) and false positives ($fp$) are then calculated for the given recall ($r$) and precision ($p$) (lines~\ref{algoline:tp_calc}-\ref{algoline:fp_calc}). The \Call{RandomChoice}{$M_x, n$} procedure returns $n$ number of randomly selected methods from the set $M_x$, where $x \in \{b,n\}$. $C_b$ is assigned a set of randomly picked $tp$ number of buggy and $fp$ number of non-buggy method indices (line~\ref{algoline:buggy_method_classfic}). $C_b$ is the set of buggy method indices as classified by the simulated defect predictor. The output is the set $C = \{c_1,\ldots,c_k\}$, where $c_i = 1$ if the method with index $i$ is labelled as buggy and $c_i = 0$ if the method with index $i$ is labelled as not buggy (line~\ref{algoline:classifications}).

\vspace{-3mm}
\subsection{Search-Based Software Testing Guided By Defect Prediction}
\label{subsec:sbst_dpg}

We incorporate buggy method predictions in DynaMOSA~\citep{panichella2017automated}, the state-of-the-art SBST technique, to guide the search for test cases towards likely buggy methods. 
DynaMOSA tackles the test generation problem as a many objective optimisation problem, where each coverage target in the program, e.g., branch and statement, is an objective to optimise. 
It is more effective at achieving high branch, statement and strong mutation coverage than previously proposed SBST techniques (\citep{fraser2012whole, rojas2017detailed, panichella2015reformulating})~\citep{panichella2017automated}. In the next sections, we refer to the DynaMOSA approach guided by the defect predictor as \textit{\SBST{}}. \SBST{} is presented in Algorithm~\ref{algo:obm}. It shares the same search steps and genetic operators as DynaMOSA, except for the updated steps shown in blue colour in Algorithm~\ref{algo:obm}. 

\SBST{} receives as input a class with methods labelled as buggy or non-buggy, which are labels that can be obtained using existing defect predictors~\citep{giger2012method, hata2012bug}. In our study, \SBST{} receives these labels from defect predictor simulations (Section~\ref{subsec:dp_sim}). 

\SBST{} devotes all the search resources to find tests that cover likely buggy methods, thereby increasing the chances of finding bugs. Initially, \SBST{} filters out the coverage targets that are deemed to not contain buggy methods as indicated by the defect prediction information, and keeps only targets that contain likely buggy methods (as shown in line~\ref{algoline_obm:filter_targets} of Algorithm~\ref{algo:obm} and described in Section~\ref{subsec:filter_targets_obm}). 

It also generates more than one test case for all the selected buggy targets, hence, further increases the chances of finding bugs (lines~\ref{algoline_obm:update_archive_rp},~\ref{algoline_obm:update_targets_rp},~\ref{algoline_obm:update_archive_op} and~\ref{algoline_obm:update_targets_op} and described in Section~\ref{subsec:dyna_tar_archive})~\citep{perera2020defect}. 

To generate more than one test case for all the likely buggy targets, \SBST{} does not remove a target once it is covered during the search. 
This is likely to cause \SBST{} to miss nontrivial targets in the search and keep on generating tests to cover more trivial targets~\citep{rojas2017detailed}. 
To address this, we use a method to dynamically disable targets from the search based on their current test coverage and number of independent paths (line~\ref{algoline_obm:switch_off_targets} and described in Section~\ref{subsec:balanced_coverage}). We refer to this as balanced test coverage. 
This ensures that the nontrivial targets have an equal chance of being covered compared to the targets that are easier to cover.

\SBST{} randomly generates a set of test cases that forms the initial population (line~\ref{algoline_obm:random_pop}). Then, it evolves this initial population through creating new test cases via crossover and mutation (line~\ref{algoline_obm:gen_offspring}), and selecting test cases to the next generation (line~\ref{algoline_obm:select_pop}), until a termination criteria, such as maximum time budget, is met.

\begin{algorithm}[!ht]
\caption{SBST Guided By Defect Prediction}\label{algo:obm}
\hspace*{\algorithmicindent} \textbf{Input:} \Comment{}\\
\hspace*{\algorithmicindent} $U = \{u_{1}, \ldots, u_{k}\}$ \Comment{the set of coverage targets of CUT}\\
\hspace*{\algorithmicindent} $G = \langle N,E \rangle$ \Comment{control dependency graph of the CUT}\\
\hspace*{\algorithmicindent} $\phi: E \rightarrow U$ \Comment{partial map between edges and targets}\\
\hspace*{\algorithmicindent} \textcolor{blue}{$C$ = \{$c_{1}, \ldots, c_{m}$\} \Comment{the set of defectiveness classifications for methods in the CUT}}
\begin{algorithmic}[1]
\Procedure{SBST}{}
\State \textcolor{blue}{$U_{B} \gets$ \Call{FilterTargets}{$U, C$}} \label{algoline_obm:filter_targets}
\State \textcolor{blue}{$L \gets$ \Call{IndependentPaths}{$G$}} \textcolor{blue}{\Comment{$L$ is a vector of the number of independent paths for each edge}} \label{algoline_obm:independent_paths}
\State $U^* \gets$ targets in \textcolor{blue}{$U_B$} with no control dependencies \label{algoline_obm:init_targets}
\State $P_{0}  \gets$ \Call{RandomPopulation}{$M$} \Comment{$M$ is the population size} \label{algoline_obm:random_pop}
\State $A \gets$ \Call{UpdateArchive}{$P_0, \emptyset, \textcolor{blue}{U_B}$} \Comment{$A$ is the archive} \label{algoline_obm:update_archive_rp}
\State $U^* \gets$ \Call{UpdateTargets}{$U^*, G, \phi, \textcolor{blue}{U_B}$} \label{algoline_obm:update_targets_rp}
\For{$r\gets0\, ;\, !$terminationCriteria; \,$r$++}
	\State $Q_{r}  \gets$ \Call{GenerateOffspring}{$P_r$} \label{algoline_obm:gen_offspring}
	\State $A \gets$ \Call{UpdateArchive}{$Q_r, A, \textcolor{blue}{U_B}$} \label{algoline_obm:update_archive_op}
	\State $U^* \gets$ \Call{UpdateTargets}{$U^*, G, \phi, \textcolor{blue}{U_B}$} \label{algoline_obm:update_targets_op}
	\State $R_r \gets$ $P_r \cup Q_r$
	\State \textcolor{blue}{$U^*  \gets$ \Call{SwitchOffTargets}{$U^*, A, L, \phi$}} \label{algoline_obm:switch_off_targets}
	\State $P_{r+1} \gets$ \Call{SelectPopulation}{$R_r, U^*, M$} \label{algoline_obm:select_pop}
\EndFor
\State $T \gets A$ \Comment{Update the final test suite $T$}
\State \Call{Return}{$T$}
\EndProcedure
\end{algorithmic}
\end{algorithm}

\vspace{-2mm}
\subsubsection{Filtering Targets with Defect Prediction}
\label{subsec:filter_targets_obm}

A defect predictor classifies the methods of the class under test (CUT) as buggy or non-buggy, denoted as $c_i$, where 
\vspace{-2mm} $$
  c_i= \begin{cases}
    1       & \quad \text{if method with index } i \text{ is predicted as buggy}\\
   0  & \quad \text{otherwise }
  \end{cases}
$$ 
This information is used to filter out the likely non-buggy targets from the set of all targets $U$ using the classifications given (line~\ref{algoline_obm:filter_targets}). 
Spending the limited search resources on covering non-buggy targets is likely to be ineffective when it comes to finding bugs. Filtering out targets that are unlikely to be buggy allows the search to focus on test cases that cover the likely buggy targets (i.e., $\forall u \in U_B$), hence, generating more effective test cases faster than other approaches which search for tests in all the targets in the CUT.

\vspace{-2mm}
\subsubsection{Dynamic Selection of Targets and Archiving Tests}
\label{subsec:dyna_tar_archive}

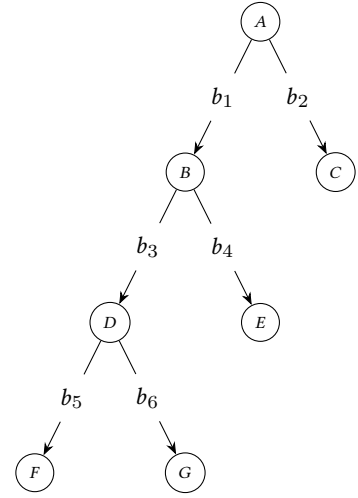
\begin{figure}[!ht]
\centering
\begin{tikzpicture}[]
  
  \node[graph vertex] (273-1) at (3,0) {\scriptsize $A$};
  \node[graph vertex] (273-2) at (2,-2) {\scriptsize $B$};
  \node[graph vertex] (276-1) at (4,-2) {\scriptsize $C$};
  \node[graph vertex] (274) at (1,-4) {\scriptsize $D$};
  \node[graph vertex] (277) at (0,-6) {\scriptsize $F$};
  \node[graph vertex] (276-2) at (3,-4) {\scriptsize $E$};
  \node[graph vertex] (279-2) at (2,-6) {\scriptsize $G$};
  
  \begin{scope}[>={Stealth[black]},
              every node/.style={fill=white,circle},
              every edge/.style={draw=black}, font=\normalsize]
\path [->] 

(273-1) edge node {$b_{1}$} (273-2)
     edge node {$b_{2}$} (276-1)
(273-2) edge node {$b_{3}$} (274)
    edge node {$b_{4}$} (276-2)
(274) edge node {$b_{5}$} (277)
     edge node {$b_{6}$} (279-2);

 \end{scope}
\end{tikzpicture}
\caption{Control Dependency Graph}
\label{fig:cdg_example}
\vspace{-4mm}
\end{figure}

There are structural dependencies of targets that should be considered when selecting objectives, i.e., targets, to optimise. For instance, some of the targets can be covered only if their control dependent targets are covered. 
To better understand this, let us consider the following example in Figure~\ref{fig:cdg_example}. 
Assume the test generation scenario is to optimise branch coverage and $b_1$, $b_2$, $b_3$, $b_4$, $b_5$ and $b_6$ are the branches to be covered. 
Branch $b_1$ holds a control dependency link to $b_3$ and $b_4$, which means that they can be covered only if $b_1$ is covered by a test case. 
If an SBST technique optimises test cases to cover $b_3$ and $b_4$, while $b_1$ is uncovered, this will unnecessarily increase the computational complexity of the algorithm because of the added objectives, i.e., $b_3$ and $b_4$, to the search without any added benefit. 
To address this, DynaMOSA dynamically selects targets to the search only when their control dependent targets are covered~\citep{panichella2017automated}. In our example, $b_3$ and $b_4$ are added to the search only when $b_1$ is covered.

At the start of the search, \SBST{} selects the set of targets $U^* \subseteq U_B$ that do not have control dependencies (line~\ref{algoline_obm:init_targets}). These are the targets \SBST{} can cover without requiring to cover any other targets in the program. At any given time in the search, it searches for test cases to cover only the targets in $U^*$. 

Once a new population of test cases is generated (lines~\ref{algoline_obm:random_pop} and~\ref{algoline_obm:gen_offspring}), the procedure \Call{UpdateTargets}{} is executed to update $U^*$ by adding new targets to the search. The procedure \Call{UpdateTargets}{} adds a target $u \in U_B$ to $U^*$ only if the control dependent targets of $u$ are covered as explained with the example above. 

\SBST{} maintains an archive of test cases found during the search which cover the selected targets. 
Once the search finishes, this archive forms the final test suite. 
Unlike in DynaMOSA, we configure the \Call{UpdateTargets}{} procedure to not remove a covered target from $U^*$ and the \Call{UpdateArchive}{} procedure (lines~\ref{algoline_obm:update_archive_rp} and \ref{algoline_obm:update_archive_op}) to archive all the test cases that cover the selected targets $u \in U_B$. 
This way, \SBST{} can generate more than one test case for each target $u \in U_B$, hence increasing the bug detection capability of the generated test suites~\citep{perera2020defect}. Perera et al.~\citep{perera2020defect} showed that DynaMOSA finds up to 79\% more bugs when it was configured to not remove covered targets from the search and retain all the generated tests.

\subsubsection{Balanced Test Coverage of Targets}
\label{subsec:balanced_coverage}

As we discussed in Section~\ref{subsec:dyna_tar_archive}, \SBST{} does not remove covered targets from $U^*$ to allow the search to find more than one test case for each target. 
While this benefits \SBST{} in terms of bug detection, one downside of this approach is that trivial targets are covered more often than they need to be. 
Hence, we propose a method to balance the test coverage among the targets in the CUT.

We use the control dependency graph (CDG) in Figure~\ref{fig:cdg_example} to explain how we balance the test coverage. Assume the goal of the test generation scenario is to maximise the branch coverage. 
There are always more test cases that cover the branches closer to the root node, e.g., $b_{1}$, than the branches closer to the leaf nodes of the CDG, e.g., $b_{5}$. 
This is because the execution of a test case can take many paths in the program once it reaches branches like $b_{1}$. For instance, there are 3 independent paths that leads from $b_{1}$ in the example. On the other hand, the execution of a test case can only take one path from $b_{5}$, i.e., the exit path. 
Therefore, to balance the test coverage of targets in the CUT, we want to ensure all targets have an equal number of tests per an independent path that leads from the respective target. 
For example, if we assume the number of tests that cover $b_{1}$ to be 90, then there should be 60 and 30 test cases that cover $b_{3}$ and $b_4$, respectively. Those 60 tests that cover $b_3$ should be equally distributed among $b_5$ and $b_6$.

\SBST{} calculates the number of independent paths of each edge in the control dependency graph $G$ of the program (line~\ref{algoline_obm:independent_paths}). 
The CDG, i.e., $G = \langle N, E \rangle$, consists of nodes $n \in N$ and edges $e \in E \subseteq N \times N$. 
The nodes represent statements in the program. 
The edges represent control dependencies between the statements. 
The procedure \Call{IndependentPaths}{} calculates the number of independent paths for each edge $e \in E$. 
When calculating the number of independent paths of an edge $e$, the \Call{IndependentPaths}{} procedure assumes the paths start at $e$, however, the actual execution of the paths start at the root node, e.g., node $A$ in Figure~\ref{fig:cdg_example}. 
In our example in Figure~\ref{fig:cdg_example}, the independent paths starting from $b_3$ are $b_3-b_5$ and $b_3-b_6$. 
Finally, all the targets that are directly control dependent by an edge $e$ have the same number of independent paths as that of $e$.

\begin{algorithm}[!ht]
\caption{Temporarily Removal of Targets to Balance Test Coverage}\label{algo:switchofftargets}
\begin{algorithmic}[1]
\Procedure{SwitchOffTargets}{$U^*, A, L, \phi$}
\State $N_P \gets$ \Call{NodesWithPredicates}{$G$} \label{algoline:nodes_predicates}
\For{$n \in N_P$}
    \State $\{e_{n,T}, e_{n,F}\} \gets$ outgoing edges in $G$ from node $n$
    \State $l_{n,T} \gets$ \Call{GetIndependentPaths}{$L, e_{n,T}$} \label{algoline:get_ip_1}
    \State $l_{n,F} \gets$ \Call{GetIndependentPaths}{$L, e_{n,F}$} \label{algoline:get_ip_2} 
    \State $u_{n,T} \gets$ \Call{RandomChoice}{$\{\phi(e_{n,T})\}$} \label{algoline:random_choice_1} 
    \State $u_{n,F} \gets$ \Call{RandomChoice}{$\{\phi(e_{n,F})\}$} \label{algoline:random_choice_2} 
    \State $A_{n,T} \gets$ \Call{GetTests}{$A, u_{n,T}$} \label{algoline:get_tests_1}
    \State $A_{n,F} \gets$ \Call{GetTests}{$A, u_{n,F}$} \label{algoline:get_tests_2}
    \If{$\dfrac{|A_{n,T}|}{l_{n,T}} > \dfrac{|A_{n,F}|}{l_{n,F}}$}
    \State $U^* \gets U^* - \{\phi(e_{n,T})\}$
    \ElsIf{$\dfrac{|A_{n,T}|}{l_{n,T}} < \dfrac{|A_{n,F}|}{l_{n,F}}$}
    \State $U^* \gets U^* - \{\phi(e_{n,F})\}$ \label{algoline:remove_u_2}
    \EndIf
\EndFor
\State \Call{Return}{$U^*$}
\EndProcedure
\end{algorithmic}
\end{algorithm}

\SBST{} dynamically switches off targets with higher test coverage from $U^*$ in every iteration to focus more on increasing the test coverage for targets which already have lower coverage (line~\ref{algoline_obm:switch_off_targets}). 
The procedure \Call{SelectPopulation}{} selects test cases to the next generation considering only these targets with low test coverage. 
Hence, this paves way for the search to find more test cases in the next generation that cover these targets and eventually make all targets to have an equitable test coverage. 
This ensures that nontrivial targets also receive a good coverage in the presence of more trivial targets.

The procedure \Call{SwitchOffTargets}{} starts with finding the set of nodes with predicates $N_P$ in $G$ (line~\ref{algoline:nodes_predicates} in Algorithm~\ref{algo:switchofftargets}). 
In our running example, $N_P = \{A, B, D\}$. 
Next, it fetches the number of independent paths from the outgoing edges of each node $n \in N_P$ (lines~\ref{algoline:get_ip_1}-\ref{algoline:get_ip_2}). 
For the node $A$, the edges $b_1$ and $b_2$ have 3 and 1 independent paths, respectively. 
We consider the test coverage is equal among all the control dependent targets of an edge, including the edge itself as well. 
Therefore, it randomly selects a control dependent target from each outgoing edge of $n$ (lines~\ref{algoline:random_choice_1}-\ref{algoline:random_choice_2}) and finds the test coverage of each edge (lines~\ref{algoline:get_tests_1}-\ref{algoline:get_tests_2}). 
In the case of a test generation scenario for maximising branch coverage, the edges $b_1$, $b_2$, etc. also become targets of the CUT. 
Then, it finds the edge which has the largest test coverage per an independent path, and removes all the control dependent targets of that edge from $U^*$ (lines~\ref{algoline:get_tests_1}-\ref{algoline:remove_u_2}). 
If we assume there are 30 and 20 tests in the archive which cover $b_1$ and $b_2$, respectively, then it removes $b_2$ from $U^*$ since $b_2$ has 20 (=20/1) tests per an independent path, while $b_1$ has only 10 (=30/3).

\vspace{-4mm}
\section{Design of Experiments}
\label{sec:design_experiments}

We design a set of experiments to evaluate the effectiveness of \SBST{} in terms of finding bugs when using defect predictors with 12 different levels of imprecision as described in Section~\ref{subsec:dp_sim} (\textit{RQ}). 
We use the bugs from the Defects4J dataset as the experimental subjects~\citep{defects4jweb} (see Section~\ref{subsec:exp_subjects}). 

To account for the randomness of the defect prediction simulation algorithm (Algorithm~\ref{algo:dp_sim}), we repeat the simulation runs 5 times for each defect predictor configuration (i.e., recall and precision pair). For each of these simulation runs, we repeat the test generation runs 5 times, to account for the randomness in \SBST{}. 

Once tests are generated and evaluated for bug detection, we conduct two-way ANOVA test to statistically analyse the effects of recall and precision of the defect predictor on the bug detection effectiveness of \SBST{}.

\vspace{-5mm}
\subsection{Experimental Subjects}
\label{subsec:exp_subjects}

We use the Defects4J dataset (version 1.5.0)~\citep{defects4jweb, just2014defects4j} as our benchmark. It contains 438 real bugs from 6 real-world open source Java projects. 
In our experiments, we remove 18 bugs altogether from the dataset; 4 deprecated bugs, 12 bugs that do not have buggy methods, and 2 bugs for which \SBST{} generated uncompilable tests (e.g., method signature is changed in the bug fix). 
Thus, we evaluate \SBST{} on a total of 420 bugs. 
The bugs are drawn from the following projects; JFreeChart (25 bugs), Closure Compiler (170 bugs), Apache commons-lang (59 bugs), Apache commons-math (104 bugs), Mockito (37 bugs), and Joda-Time (25 bugs).

The Defects4J benchmark gives a buggy version and a fixed version of the program for each bug in the dataset. 
The fixed version is different to the buggy version by the applied patch to fix the bug, which indicates the location of the bug. 
We label all the methods that are either modified or removed in the bug fix as buggy methods~\citep{sohn2019empirical}.

Defects4J is widely used for research on automated unit test generation~\citep{shamshiri2015automatically, perera2020defect, gay2017fitness}, automated program repair~\citep{aleti2020apr}, fault localisation~\citep{pearson2017evaluating}, test case prioritisation~\citep{paterson2019empirical}, etc. 
This makes Defects4J a suitable benchmark for evaluating \SBST{}, as it allows us to compare our results to existing work.

\vspace{-5mm}
\subsection{Prototype}

DynaMOSA is implemented in the state-of-the-art SBST tool, EvoSuite~\citep{fraser2011evolutionary}. 
EvoSuite is an automated test generation framework that generates JUnit test suites for java programs~\citep{evosuitegithub, evosuiteweb}. 
EvoSuite is actively maintained and evaluated for its effectiveness in terms of bug finding on both industrial and open source projects~\citep{shamshiri2015automatically,perera2020defect,almasi2017industrial,gay2017fitness}. 
For the experimental evaluation, we implement the changes described in Section~\ref{subsec:sbst_dpg} for \SBST{}. 
The changes are implemented within EvoSuite version 1.0.7, forked from the GitHub repository~\citep{evosuitegithub} on June 18$^{th}$, 2019. 
We also implement the defect predictor simulator as described in Section~\ref{subsec:dp_sim}. 
The prototypes are available to download from here: https://figshare.com/s/a8d75f161b8cfa11d297

\vspace{-4mm}
\subsection{Parameter Settings}
\label{sec:parameter_settings}

We use the default parameter settings of EvoSuite~\citep{fraser2012whole} and DynaMOSA~\citep{panichella2017automated} except for the parameters mentioned in the next paragraphs. 
Parameter tuning of SBST techniques is a long and expensive process~\citep{arcuri2013parameter}. According to Arcuri and Fraser~\citep{arcuri2013parameter}, EvoSuite with default parameter values performs on par compared to EvoSuite with tuned parameters.

\textit{Time Budget:} We set 2 minutes as time budget per CUT for test generation. In practice, the time budget allocated for SBST tools depends on the size of the project, frequency of test generation runs and availability of computational resources in the organisation. 

Real world projects are usually very large and can have thousands of classes~\citep{broy2007engineering}. If an SBST tool runs test generation for 2 minutes per class, then it will take at least 33 hours to finish the task for the whole project. 

To address this issue, practitioners can adapt the SBST tools in their continuous integration (CI) systems~\citep{fowler2006continuous}. 
However, the introduction of new SBST tools to the CI system should not make the existing processes in the system idle~\citep{perera2020defect}. 

Thus, given the limited computational resources available in practice~\citep{campos2014continuous} and the expectation of faster feedback cycles from testing in agile development prompt the necessity of frequent test generation runs with limited testing budget. 
Therefore, we decide that 2 minutes per class is a reasonable time budget in a usual resource constrained environment.

\textit{Coverage criteria:} We use branch coverage as coverage criterion in \SBST{}. 
EvoSuite is more effective in terms of finding bugs when it is using branch coverage as the coverage criterion compared to other single criteria~\citep{gay2017generating}. 
According to Gay~\citep{gay2017generating}, some criteria combinations perform better than branch coverage. However, there were other combinations that perform worse than branch coverage. Hence, they did not recommend a strategy to combine criteria. Therefore, we decide to use the most effective single criterion, i.e., branch coverage, in our experimental evaluation. 

\textit{Termination criteria:} We use only the maximum time budget as the termination criterion. Stopping the search after it covers all the targets is detrimental to bug detection~\citep{perera2020defect}. The search needs to utilise the full time budget to generate as many tests for each target in the CUT in order to increase the chances of detecting bugs. Therefore, we terminate the search for test cases only when the allocated time budget runs out.

\textit{Test suite minimisation:} We disable test suite minimisation since all the test cases in the archive form the final test suite (see Section~\ref{subsec:dyna_tar_archive}).

\textit{Assertion strategy:} We choose all possible assertions as the assertion strategy because the mutation-based assertion filtering can be computationally expensive and can lead to timeouts~\citep{shamshiri2015automatically,perera2020defect}.

\vspace{-3.5mm}
\subsection{Experimental Protocol}

We run experiments with \SBST{} using defect predictors with 12 different levels of imprecision as described in Section~\ref{subsec:dp_sim}. 
For each bug in the Defects4J dataset, we checkout the buggy version of the project and collect the ground truth labels for the buggy and non-buggy methods. 
If a method is either modified or removed in the bug fix, we label that method as a buggy method, and non-buggy otherwise~\citep{sohn2019empirical}. 
Then, for each of the six projects in the dataset, we combine the ground truth labels from all the bugs respective to each project. 
For example, we combine the labels from all the 104 bugs from Apache commons-math project. 
Then, we simulate defect prediction outcomes for each project using the defect prediction algorithm described in Section~\ref{subsec:dp_sim}. 

We assume an application scenario of generating tests to find bugs not only limited to regressions, but also the bugs introduced to the code in various times in development. 
Therefore, we run test generation on the buggy version of the projects. 
We measure the bug finding effectiveness of \SBST{} only on the Defects4J bugs. 
Thus, we only run test generation for buggy classes, i.e., classes that are modified in the bug fixes, in the projects.

For each level of defect predictor imprecision, we run test generation with \SBST{} 25 times for each bug in the dataset. 
Consequently, we have to run a total of 12 (levels of defect prediction imprecision) $*$ 25 (repetitions) $*$ 482 (buggy classes) $=$ 144,600 test generations.

Defects4J~\citep{defects4jweb} allows us to evaluate if the 144,600 generated test suites in the experiments find the bugs. 
First, we remove the flaky test cases in test suites using the `fix test suite' interface~\citep{defects4jweb} in Defects4J as described in~\citep{shamshiri2015automatically}. 
We use the `run bug detection' interface~\citep{defects4jweb}, which executes a test suite against the buggy and fixed versions of a program and determines if the test suite finds the bug by checking if the test execution results are different between the two versions. 
EvoSuite generates assertions assuming the program under test is correct, therefore, the generated tests should always pass when they are run against the buggy version. 
A test suite is considered broken if it is not compilable or fails when run against the buggy version of the program. 
The test suite is considered as it has missed detecting the bug if it produces the same execution results when run against the buggy and fixed versions of the program, and it is considered as it has detected the bug if the test results are different.

\vspace{-2mm}
\section{Results}

We present the results for our research question following the method described in Section~\ref{sec:design_experiments}. Our aim is to evaluate the effectiveness of bug finding performance of \SBST{} when using imprecise defect predictors.

\vspace{-2mm}
\subsection*{RQ. What is the impact of the imprecision of defect prediction on bug detection performance of SBST?}

\begin{figure}[h]
    \centering
    \includegraphics[width=0.46\textwidth]{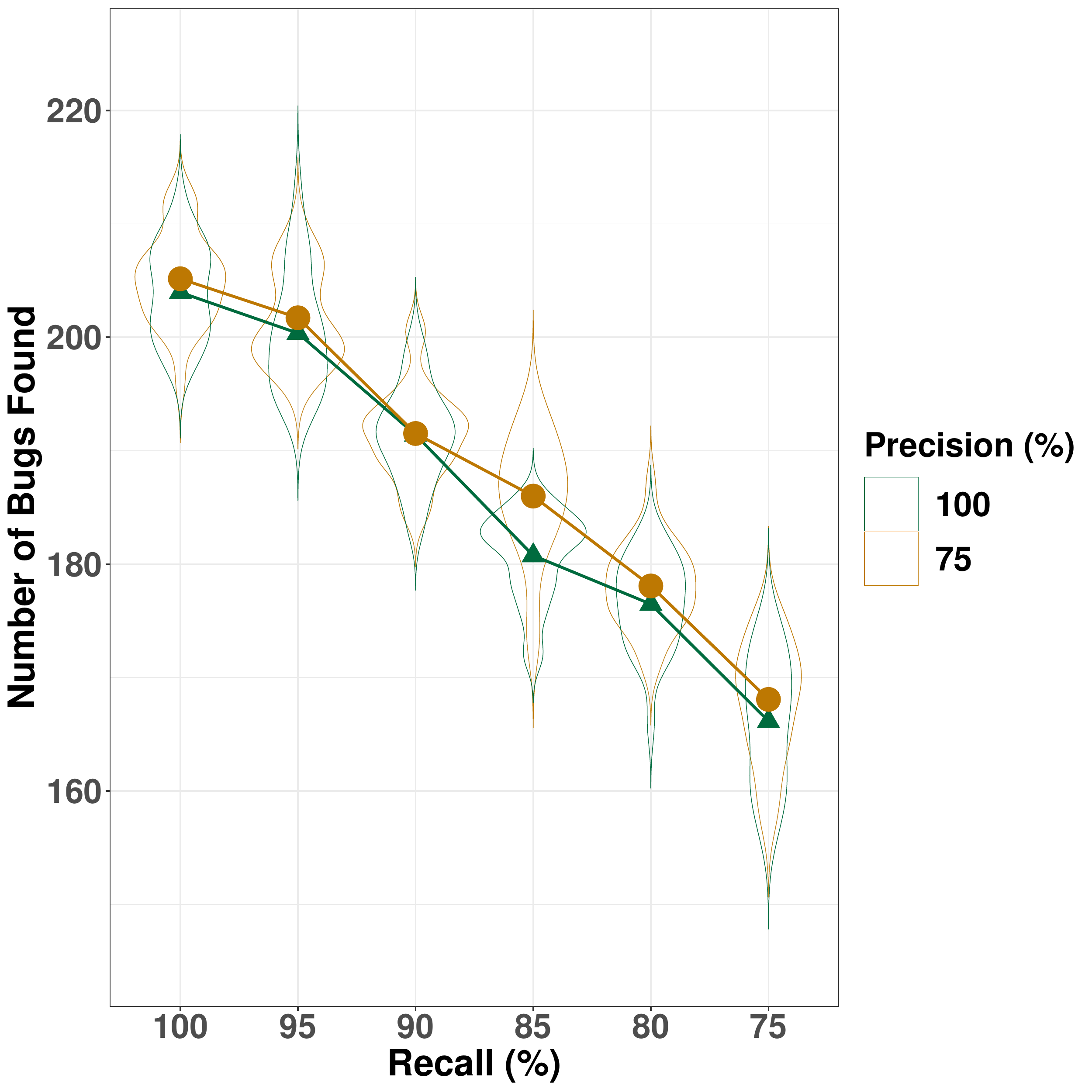}
    \caption{Distributions of the number of bugs found by \SBST{} as violin plots together with the profile plot of mean number of bugs found by \SBST{} for each combination of the groups of recall and precision.}
    \label{fig:profile_plot}
\end{figure}

Figure~\ref{fig:profile_plot} shows the distributions of the number of bugs found by \SBST{} as violin plots and the profile plot of the mean number of bugs found by \SBST{} for each combination of the factors of six recalls and two precisions. 
The two lines in our profile plot run almost parallel to each other, i.e., the two lines do not cross each other at any point. This means that there is no observable interaction effect between recall and precision. 

The two lines descent steeply from recall 100\% to 75\%. This shows that recall has an effect on number of bugs found by \SBST{}. In particular, bug detection effectiveness decreases as recall decreases.

The precision=75\% line closely follows the precision=100\% line while staying slightly above the latter, except at recall=85\%, where there is a considerable gap between the two. 
We can soon see if this difference is significant from the two-way ANOVA test results. 

\vspace{-3mm}
To statistically test the effect of each of the metrics, recall and precision, and their interaction on the number of bugs found by \SBST{}, we conduct the two-way ANOVA test. 
Prior to conducting two-way ANOVA test, we have to make sure that our data holds the following assumptions of the test. 

\begin{enumerate}
  \item The dependent variable should approximately follow a normal distribution for all the combinations of groups of the two independent variables.
  \item Homogeneity of variances exists for all the combinations of groups of the two independent variables.
\end{enumerate}

To check the first assumption, we conduct the Kolmogorov-Smirnov test~\citep{massey1951kolmogorov} for normality of the distributions ($\alpha = 0.05$) of the number of bugs found for each combination of the groups of recall and precision. Based on the results of the tests, we cannot reject our null hypothesis (p-values $\geq 0.131$), i.e., H$_0$ = the number of bugs found is normally distributed, hence we assume all the samples come from a normal distribution (i.e., H$_0$ is true).

To check the second assumption, we conduct the Bartlett’s test for homogeneity of variances ($\alpha = 0.05$) in each combination of the groups of recall and precision. Based on the results of the test, we cannot reject our null hypothesis (p-value $= 0.305$), i.e., H$_0$ = variances of the number of bugs found are equal across all combinations of the groups, hence we assume the variances are equal across all samples (i.e., H$_0$ is true).

\begin{table}[h]
\centering
\renewcommand{\arraystretch}{1.1}
\begin{tabular}{l|r|r|r|r|r}
\hline
                 & \multicolumn{1}{c|}{Df} & \multicolumn{1}{c|}{Sum Sq} & \multicolumn{1}{c|}{Mean Sq} & \multicolumn{1}{c|}{F value} & \multicolumn{1}{c}{p-value}   \\ \hline
Recall           & 5                       & 51341                       & 10268                        & 497.42                       & \textless 0.001 \\ 
Precision        & 1                       & 273                         & 273                          & 13.21                        & \textless 0.001 \\ 
Recall:Precision & 5                       & 190                         & 38                           & 1.84                         & 0.105                          \\ \hline 
Residuals        & 288                     & 5945                        & 21                        \\ \cline{1-4}
\end{tabular}
\vspace{3mm}
\caption{Summary of the two-way ANOVA test results. Df = degrees of freedom, Sum Sq = sum of squares and Mean sq = mean sum of squares.}
\label{table:two-way_ANOVA}
\end{table}

\vspace{-8mm}
Table~\ref{table:two-way_ANOVA} shows the summary of the two-way ANOVA test results. 
According to the two-way ANOVA test, recall and precision explain a significant amount of variation in number of bugs found by \SBST{} (p-values $< 0.001$). 
The test also indicates that we cannot reject the null hypothesis that there is no interaction effect between recall and precision on number of bugs found (p-value $= 0.105$). That means we can assume the effect of recall on number of bugs found does not depend on the effect of precision, and vice versa. 

To check if the observed differences among the groups are of practical significance, we measure the epsilon squared effect size ($\widehat{\epsilon}^2$)~\citep{yigit2018effect} of the variations in number of bugs found with respect to recall and precision. 
We find that the effect of recall on bug detection effectiveness is large with an effect size of 0.89, while the effect of precision is very small ($\widehat{\epsilon}^2 = 0.004$)~\citep{cohen1992power}, which can be seen from the overlapping distributions in the violin plots in Figure~\ref{fig:profile_plot} as well.

To further analyse which groups are significantly different from each other, we conduct the Tukey’s Honestly-Significant-Difference test~\citep{tukey1949comparing}. The Tukey post-hoc test shows that the number bugs found by \SBST{} is significantly different between each of the six levels of recall (p-values $< 0.002$). 
The Cohen's $d$ effect sizes of the differences between the groups of recall range from medium ($d=0.77$ for recall 95\% and 100\%) to large ($d \geq 1.33$ for all other pairs of groups).

\begin{center}
\begin{tcolorbox}[colback=black!5!white,colframe=black!75!black]
In summary, the imprecision of the defect predictor has a significant impact on the bug finding performance of SBST. 
In particular, when the recall of the defect predictor decreases, the bug detection effectiveness significantly decreases with a large effect size. 
On the other hand, we conclude that there is no meaningful practical effect of precision on the bug detection performance of SBST, as indicated by a very small effect size. 
\end{tcolorbox}
\end{center}

\subsection{Sensitivity to the Recall of the Defect Predictor}
\label{sec:sensitivity_recall}

As shown in Figure~\ref{fig:profile_plot}, \SBST{} finds less number of bugs when using defect predictors with a lower recall compared to using one with a higher recall. 
In particular, \SBST{} finds 7.5 less bugs and misses test generation for 15 bugs on average (out of 420) when the recall decreases by 5\% in our experiments. 
\SBST{} completely trusts the defect predictor and only generates tests for classes having at least one method predicted as buggy (e.g., true positive). 
The number of true positives by the defect predictor decreases when the recall decreases. 
This results in \SBST{} generating tests for a fewer number of classes as the recall decreases, hence finding less number of bugs when recall drops from 100\% to 75\%.

We identify this as a weakness of SBST when using defect predictions. 
To mitigate this, SBST techniques have to take potential errors in the predictions into account. 
One way to do this is to always generate tests for methods that are predicted buggy, while also generating tests for predicted non-buggy methods at least with a minimum probability. 
This way the SBST technique gets a chance to search for tests in incorrectly classified buggy methods (when recall \textless 100\%), while also giving higher priority to methods that are predicted buggy by the defect predictor.

\subsection{Number of Buggy Methods}

As we discussed previously, when the recall of the defect predictor decreases, \SBST{} completely misses test generation for certain bugs, hence leads to poorer bug detection. 
In our experiments, \SBST{} misses test generation for 18.2\% of the bugs on average when recall decreases from 100\% to 75\%. 
Further analysis of the results indicates that \SBST{} only misses test generation for 4.5\% of the bugs on average for the bugs that spread across multiple methods, whereas it misses 24.7\% of the bugs on average for the bugs that are concentrated into only one method. 
This suggests that the bugs that are found within only one method are more prone to the impact of recall compared to bugs that are spread across multiple methods.

To understand the effects of recall on finding bugs which are found within only one method and spread across multiple methods, we conduct Welch ANOVA test~\citep{liu2015comparing} separately for the two subsets of our dataset, i.e., bugs having only one buggy method and bugs having more than one buggy method. 
The reason for carrying out Welch ANOVA test is because our data fails the assumption of homogeneity of variances for each combination of the groups of recall for bugs having only one buggy method.

\begin{table}[h]
\centering
\renewcommand{\arraystretch}{1.1}
\begin{tabular}{l|r|r|r|r}
\hline
                 & \multicolumn{1}{c|}{Num Df} & \multicolumn{1}{c|}{Denom Df} &  \multicolumn{1}{c|}{F value} & \multicolumn{1}{c}{p-value}   \\ \hline
\# buggy methods $>1$           & 5.00                       & 137.06                                              & 67.24                       & \textless 0.001 \\
\# buggy methods $=1$        & 5.00                       & 136.68                                                  & 395.91                        & \textless 0.001 \\ \hline
\end{tabular}
\vspace{3mm}
\caption{Summary of the Welch ANOVA test results. Num Df = degrees of freedom of the numerator and Denom Df = degrees of freedom of the denominator.}
\vspace{-5mm}
\label{table:welch_ANOVA}
\end{table}

\begin{figure}[h]
    \centering
    \includegraphics[width=0.34\textwidth]{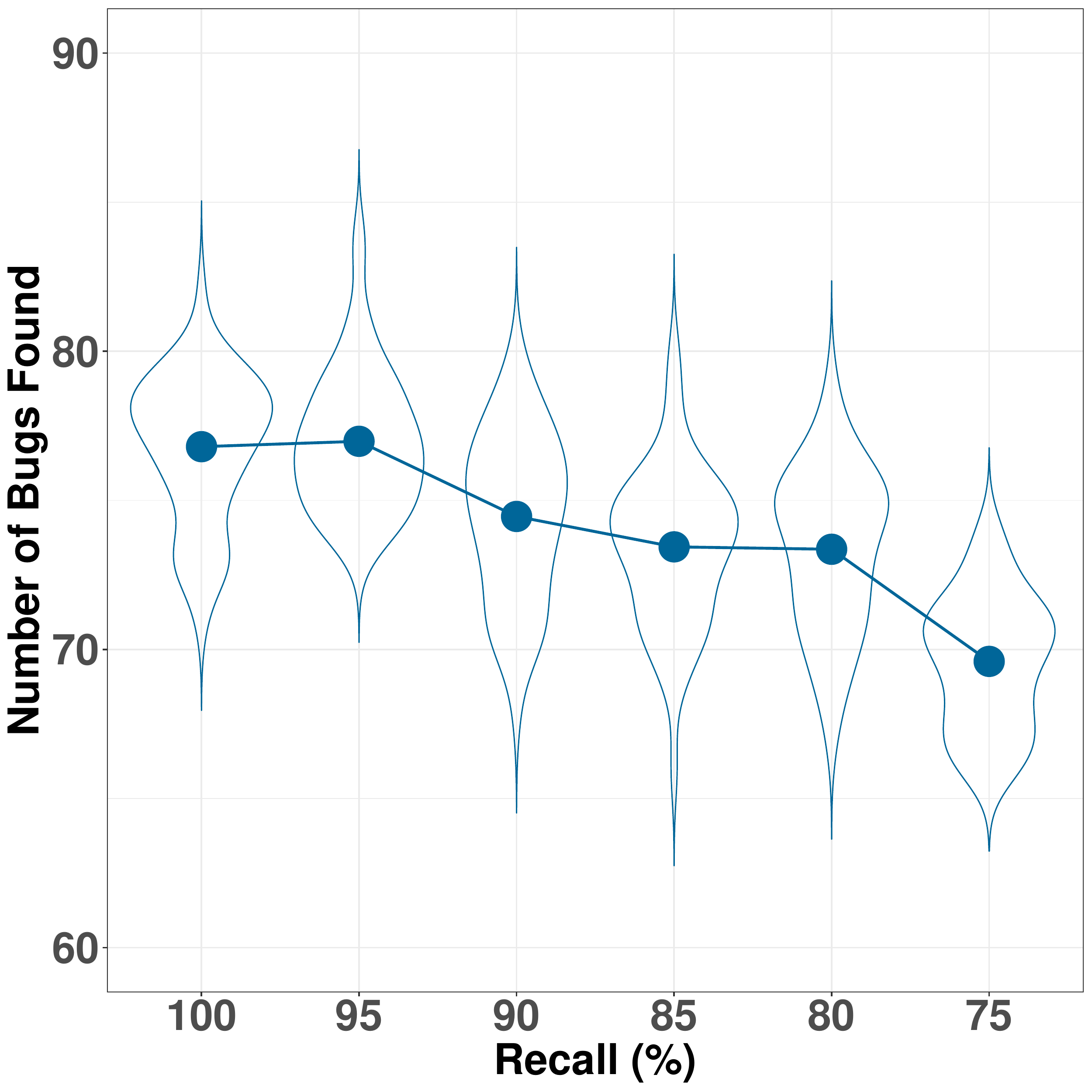}
    \caption{Distributions of the number of bugs found by \SBST{} as violin plots together with the means plot of number of bugs found by \SBST{} for the groups of recall. Only for the bugs that have more than one buggy method. Total number of bugs = 135.}
    \vspace{-4mm}
    \label{fig:means_plot_bm_ge_2}
\end{figure}

The results of the Welch ANOVA test are shown in Table~\ref{table:welch_ANOVA}. 
There are 135 bugs which have more than one buggy method. 
The results for these bugs show that overall recall has a significant effect on number of bugs found by \SBST{} (p-value \textless 0.001) with a large effect size ($\widehat{\epsilon}^2 = 0.53$)~\citep{carroll1975sampling}. 
However, the Games-Howell post-hoc test reveals that the bug detection effectiveness is not significantly different between recall 80\%, 85\% and 90\%, and 95\% and 100\%. 
This can be seen in the violin plots in Figure~\ref{fig:means_plot_bm_ge_2} as well. 

\begin{figure}[h]
    \centering
    \includegraphics[width=0.34\textwidth]{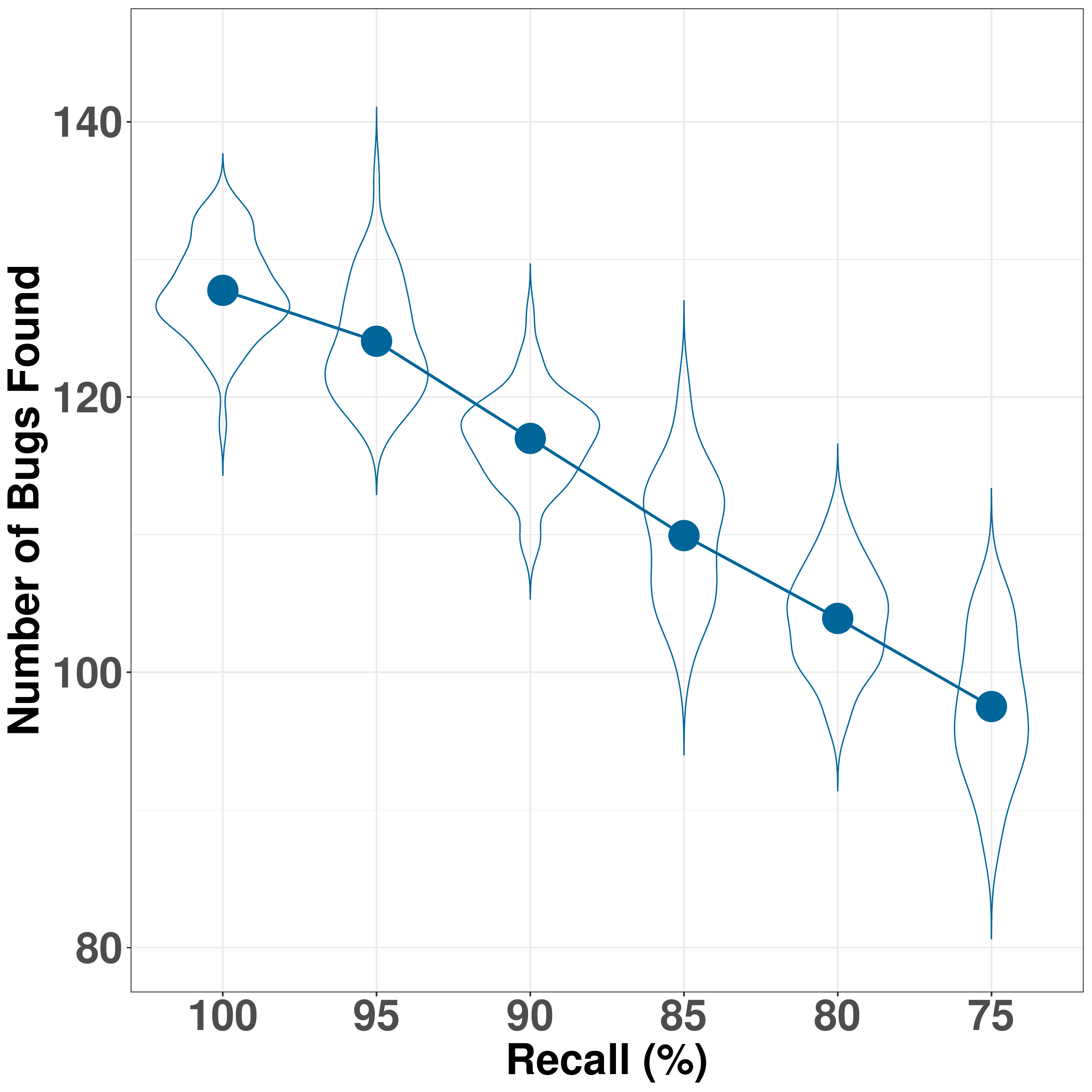}
    \caption{Distributions of the number of bugs found by \SBST{} as violin plots together with the means plot of number of bugs found by \SBST{} for the groups of recall. Only for the bugs that have one buggy method. Total number of bugs = 285.}
    \vspace{-3mm}
    \label{fig:means_plot_bm_le_1}
\end{figure}

There are 285 bugs which have only one buggy method. 
The results of Welch ANOVA test for these bugs show that recall has a significant effect on number of bugs found by \SBST{} (p-value \textless 0.001) with a large effect size ($\widehat{\epsilon}^2 = 0.87$). 
The Games-Howell post-hoc test confirms that the number of bugs found by \SBST{} is significantly different between each group of recall (p-values \textless 0.001) with large effect sizes ($d \geq 0.98$) as can be seen in Figure~\ref{fig:means_plot_bm_le_1}.

In summary, we find that recall has a significant effect on bug detection effectiveness of \SBST{} regardless of whether the bugs are found within one method or spread across multiple methods. 
However, for the bugs that are spread across multiple methods, the effect size of recall effect is smaller when compared to bugs that are found within one method ($0.53 < 0.87$). 
In contrast to bugs that are found within one method, the effect of recall is not significant between the groups of recall 80\%, 85\% and 90\%, and 95\% and 100\% for the bugs that are spread across multiple methods. 

\vspace{-3mm}
\subsection{Sensitivity to the Precision of the Defect Predictor}

According to the two-way ANOVA test, the precision of the defect predictor has a statistically significant effect, although with a very small effect size, which suggests the effect is not of meaningful practical significance. 
Precision is associated with false positives (Equation~\eqref{eq:precision}), i.e., non-buggy methods predicted as buggy by the defect predictor. 
Change of precision from 100\% to 75\% means that there are false positives in the defect prediction results. 
We investigate the buggy method labels produced by the defect predictor and the bug finding results of \SBST{} in our experiments to find out if false positives have actually helped \SBST{} to find more bugs. 
We find that the false positives have not contributed to the bug finding performance of \SBST{}. 
We conclude that the impact of precision is not of practical significance to the bug finding performance of SBST.

\vspace{-3mm}
\section{Discussion}

Defect predictors have mainly been used to provide a list of likely defective parts of a program (e.g., classes and methods) to programmers, who then manually inspect or test the likely defective parts to find the bugs~\citep{lewis2013does, dam2019lessons}. 
In this context, the precision of the defect predictor is very important~\citep{wan2018perceptions}. 
Poor precision of the defect predictor means there are higher false positives. 
Higher false positives can waste developers' time and lead to losing their trust on the prediction results~\citep{lewis2013does}. 
However, when the defect predictions are consumed by another automated testing technique such as SBST, this may not be the case. 
In the context of SBST, our study reveals contrasting findings. 
We find that the effect of precision on the bug detection performance of SBST is negligible, while the recall of the predictor has a significant impact with a large effect size.

We recommend that programmers improve the recall of the defect predictor at the cost of precision to achieve good performance in SBST guided by defect prediction. 
There is a trade-off between recall and precision of a defect predictor~\citep{koru2005building}. 
Defect predictors are usually good at detecting bugs (i.e., high recall) at the expense of false positives (i.e., low precision). 
Our study shows the bug detection effectiveness of \SBST{} is highly sensitive to recall, while the effect of precision is negligible. This means that most defect predictors proposed in the literature would be suitable for guiding SBST. 
As the scope of our study is to analyse the impact of precision in the range of an acceptable defect predictor, i.e., precision $\geq 75\%$, we cannot make any conclusions about defect predictors with precision below 75\%. 
Therefore, we can conclude that it is beneficial to increase the recall of the defect predictor by sacrificing precision as long as it is above 75\%.

\vspace{-3mm}
\section{Threats to Validity}

\textbf{Construct Validity.} To systematically investigate the impact of defect prediction imprecision, we simulate the predictions by assuming a uniform distribution. 
This means in our simulations, every method has an equal chance of being labelled as buggy or non-buggy. 
However, real defect predictors may have different distributions of their predictions depending on the underlying characteristics and nature of the prediction problem, which may impact the realism of a simulated defect predictor. 
Nevertheless, in the absence of prior knowledge about defect prediction distributions, it is reasonable to assume a uniform distribution of predictions in the defect prediction simulation. 

\SBST{} generates more than one test case for each target in the CUT. 
This increases the chances of finding bugs at the cost of larger test suites. 
Larger test suites are associated with a higher number of assertions in the tests generated by EvoSuite, which need to be manually adapted by developers in practice. 
We design our study to investigate the impact of imprecision in defect prediction on SBST along one dimension, that is the number of bugs found by the generated test suites. 
We identify investigating the impact of defect prediction imprecision on SBST in terms of the cost of manual adaptation of the generated assertions as future work, which will complement the findings of our study.

\textbf{Internal Validity.} To account for the randomness in the defect prediction simulation, we repeat the simulations 5 times for each combination of the groups of recall and precision. 
For each simulation, we repeat the test generation 5 times to account for the non-deterministic behaviour of \SBST{}. In total, we conduct 25 test generation runs for each bug and for each level of defect prediction imprecision.

\textbf{Conclusion Validity.} To account for any threats to the conclusion validity, we derive conclusions from the experimental results after conducting sound statistical tests; two-way ANOVA test, epsilon squared effect size, Tukey's Honestly-Significant-Difference test, Cohen's \textit{d} effect size, Welch ANOVA test and Games-Howell post-hoc test.

\textbf{External Validity.} We use 420 real bugs from Defects4J dataset as the experimental subjects. 
They are drawn from 6 open source projects. 
At the time of writing this paper, another 401 bugs from 11 projects were added to the Defects4J dataset. 
However, we understand that these projects do not represent all program characteristics, especially in industrial projects. 
Nevertheless, Defects4J dataset has been widely used in previous work as a benchmark~\citep{shamshiri2015automatically, perera2020defect, paterson2019empirical, aleti2020apr}. 
Future work needs to be done on investigating the impact of imprecision of defect prediction on SBST with respect to other bug datasets.

\SBST{} uses defect prediction information at method level. Our findings may not be generalised to previous work~\citep{perera2020defect, hershkovich2019prediction} which use defect prediction at a different level of granularity (class level). Nevertheless, the findings from our study will help to further explore the opportunities of combining defect predictions and SBST.

We investigate the impact of defect prediction imprecision only in the range of 75\% to 100\% for recall and precision. Therefore, our findings may not be generalised to the defect predictors which have recall or precision less than 75\%. While this choice of performance sampling in our simulation is a threat to external validity, it is also a threat to construct validity for lack of characterising all possible defect predictors. However, we opted to use this range with the justification that this is the range for an acceptable performance for a defect predictor as recommended by Zimmermann et al.~\citep{zimmermann2009cross}. 

\section{Related Work}

\subsection{Defect Prediction in Software Testing}

Defect prediction was originally proposed to provide a list of likely defective parts of a program to assist developers in code reviews~\citep{lewis2013does,googledefect}, manual testing~\citep{dam2019lessons}, etc. 
More recently, defect predictors have been used to inform automated testing techniques as well. 
G-clef~\citep{paterson2019empirical} is a test prioritisation strategy that uses the likelihood of the defectiveness of classes to prioritise test cases and it was shown to be effective at reducing the number of test cases required to find bugs. 
FLUCCS~\citep{sohn2019empirical} is a fault localisation approach that leverages the likelihood of methods being defective and it was shown to significantly outperform the state-of-the-art spectrum based fault localisation (SBFL) techniques. 
Perera et al.~\citep{perera2020defect} and Hershkovich et al.~\citep{hershkovich2019prediction} used defect predictions at class level to determine the time budget allocated to classes in a project to run test generation with SBST techniques. 
A highly likely to be defective class according to the defect predictor has more chance of being selected to run test generation~\citep{hershkovich2019prediction} or allocated a higher time budget~\citep{perera2020defect}. 
Despite showing the improved bug detection performance of the proposed SBST techniques, we find that the defect predictors used in these two works have relatively high performance, e.g., 85\% recall in~\citep{perera2020defect} and 0.95 AUC in~\citep{hershkovich2019prediction}, which can be difficult to achieve for a defect predictor. 
For example, Zimmermann et al.~\citep{zimmermann2009cross} found that only 21 out of 622 cross-project defect predictor combinations to have recall, precision and accuracy greater than 75\%. In their systematic literature review, Hall et al.~\citep{hall2011systematic} reported defect predictor performances in the ranges of 5\%-95\% and 25\%-85\% for precision and recall, respectively. 
This leads to the question of how does the variation in defect prediction performance affect the bug detection effectiveness of SBST techniques that incorporate defect prediction information. 
To address this gap, we study the impact of imprecision in defect predictions on the bug detection performance of SBST.

\subsection{Search-Based Software Testing}

Search-based software testing techniques use search algorithms like genetic algorithms to search for test cases to meet a given criteria like branch coverage~\citep{fraser2011evolutionary}. 
The test generation problem can be formulated in two ways; i) single objective formulation~\citep{rojas2017detailed, fraser2011evolutionary} and ii) many objective formulation~\citep{panichella2017automated, panichella2015reformulating}. 
In many objective optimisation, such as MOSA~\citep{panichella2015reformulating} and DynaMOSA~\citep{panichella2017automated}, SBST techniques aim to find a set of non-dominated test cases that minimise the fitness functions for all the test targets, e.g., branches.
In single objective optimisation, SBST techniques optimise whole test suites to minimise a single fitness function which is created by aggregating all the individual test target distances. A target distance measures how far away the test suite is from covering that target~\citep{fraser2011evolutionary}. 
Whole test suite generation (WS)~\citep{fraser2011evolutionary} and archive-based WS (WSA)~\citep{rojas2017detailed} are two examples for techniques that use single objective optimisation. 
Previous work showed that DynaMOSA, a state-of-the-art many objective optimisation technique, is better than single objective optimisation techniques in terms of achieving high code coverage~\citep{panichella2017automated}. 
In this paper, we study the effect of defect prediction imprecision on bug detection performance of an SBST technique that uses many objective optimisation.

\subsection{Imprecision in Defect Predictors}

There is a plethora of defect predictors which have been proposed over the past 40 years~\citep{wan2018perceptions}. 
Measures such as recall, precision, f-measure, AUC, Matthews correlation coefficient (MCC)~\citep{yao2020assessing}, etc. have been used to measure the predictive power of the defect predictors~\citep{hall2011systematic}. 
Out of these measures, recall and precision have been widely used in previous work~\citep{hall2011systematic, hosseini2017systematic} and are often preferred by practitioners~\citep{wan2018perceptions}. 
Existing defect predictors have wavering performance. 
For example, Hall et al.~\citep{hall2011systematic} reported defect predictor performances from as low as 5\% and 25\% to as high as 95\% and 85\% for precision and recall, respectively. 
Hosseini et al.~\citep{hosseini2017systematic} also reported similar findings in their systematic literature review of cross-project defect predictors. 
It is thus important to study the impact of the wavering defect prediction performance on the bug detection performance of SBST. 
In our study, we consider the recall and the precision should be greater than 75\% to be considered acceptable as recommended by Zimmermann et al.~\citep{zimmermann2009cross}, and simulate defect predictions in the range from 75\% to 100\% for recall and precision. 

Previous work report the developers' opinions about the defect predictor performance~\citep{dam2019lessons,lewis2013does,wan2018perceptions}, showing that false positives cause developers to waste their precious time on inspecting non-buggy code, which eventually leads to loosing trust on the defect predictor~\citep{dam2019lessons,lewis2013does}. 
In the eyes of the developers, higher precision is more important compared to higher recall in a defect predictor, because higher precision means low false positives~\citep{wan2018perceptions}. 
In the context of using defect prediction to guide SBST, our study reveals contrasting findings. 
In particular, precision has a negligible impact on the bug detection performance of SBST, while the effect of recall is significant.

\section{Conclusion}

We study the impact of imprecision in defect prediction on the bug detection performance of SBST. 
We use simulated defect predictors to systematically sample defect predictors in the range of 75\% to 100\% for recall and precision. 
We use the state-of-the-art SBST technique, DynaMOSA, and incorporate predictions about buggy methods as given by the simulated defect predictor to guide the search for test cases towards likely buggy methods. 
Through a comprehensive experimental evaluation on 420 bugs from the Defects4J dataset, we find that the recall of the defect predictor has a significant impact on the bug detection effectiveness of SBST with a large effect size. 
On the other hand, the impact of precision is not of meaningful practical significance as indicated by a very small effect size. 
Further analysis of the results shows that the impact of the recall for the bugs that are spread across multiple methods is smaller compared to the bugs that are found within only one method.

Based on the results of our study, we make the following recommendations: 
\begin{enumerate}
    \item SBST techniques must take into account potential errors in the predictions, especially the false negatives. 
    One way to do this is to prioritise the likely buggy parts of the program, while guiding the search towards the likely non-buggy parts with at least a minimum probability. 
    \item In the context of SBST, it is beneficial to increase the recall of the defect predictor at the expense of precision as long as it stays above 75\%. One straightforward method to do this is to lower the cut-off point of the classifier such that more components will be labelled as buggy at the cost of more false positives. 
\end{enumerate}

We identify the following directions as future work to extend this study; i) investigate the impact of defect prediction imprecision on SBST in terms of the cost of manual adaptation of generated assertions, ii) validate the findings against other bug datasets~\citep{DBLP:journals/corr/abs-2011-06244, Madeiral2019, saha2018bugs}, and iii) explore the options for using the likelihood of defectiveness of methods to guide SBST techniques.

\bibliographystyle{IEEEtranN}
\bibliography{references}

\clearpage

\end{document}